\documentclass[final,5p,times]{elsarticle}

\usepackage{tikz}
\usetikzlibrary{positioning}

\usepackage{caption}
\usepackage{csquotes}
\usepackage{amssymb}
\usepackage{amsmath}
\usepackage{amsthm}
\usepackage{float}
\usepackage{algorithm}
\usepackage{algpseudocode}
\usepackage[colorlinks=true,linkcolor=blue,citecolor=blue,urlcolor=blue]{hyperref}

\journal{Information Processing Letters}

\newcommand{\set}[1]{\left\{ #1 \right\}}
\newcommand{\code}[1]{\texttt{\detokenize{#1}}}
\makeatletter
\newcommand{\breaktwo}{\if@twocolumn\\\fi}
\makeatother
\begin{document}

\begin{frontmatter}

\title{Global Predecessor Indexing: Avoiding Binary Search in Weighted Job Scheduling}

\author{
Amit Joshi \\
\href{mailto:amitjoshi2724@gmail.com}{\code{amitjoshi2724@gmail.com}} \\
\href{mailto:amit.joshiusa@gmail.com}{\code{amit.joshiusa@gmail.com}}
}
\makeatletter
\if@twocolumn
  \address{Independent Researcher\vspace{-5em}}
\else
  \address{Independent Researcher\vspace{-4em}}
\fi
\makeatother
\begin{abstract}
We present an improved solution to the Weighted Job Scheduling (WJS) problem. While the classical dynamic programming (DP) solution for $n$ jobs runs in $O(n \log(n))$ time due to comparison-based sorting and per-job binary search \cite{kleinberg2005}, we eliminate the binary search bottleneck. In its place, we introduce a novel multi-phase preprocessing technique called \emph{Global Predecessor Indexing (GPI)}, which computes the latest non-overlapping job (i.e., the predecessor) for all jobs via a two-pointer linear-time pass after sorting. This yields a time complexity of $O(S(n) + n)$ where $S(n)$ is the time to sort all jobs. GPI enables direct use in the classical DP recurrence. When combined with linear-time sorting, GPI yields a complete $O(n)$ solution. Even with comparison-based sorting, GPI significantly outperforms the classical solution in practice by avoiding repeated binary searches in favor of the more cache-efficient extra sort and two-pointer pass.
\end{abstract}

\end{frontmatter}
\section{Introduction}
\label{sec:intro}

The Weighted Job Scheduling (WJS) or Weighted Interval Scheduling (WIS) problem is a classic combinatorial optimization task where each job has a time interval and weight (or value), and the goal is to select a subset of non-overlapping jobs that maximizes total weight. This problem arises in compiler register allocation, where weighted intervals represent variable lifetimes; the goal is to assign variables to registers without conflicts and prioritize the most critical variables. It also appears in server backup scheduling, where jobs have time windows and priority weights; the objective is to select non-overlapping jobs that maximize the total priority-weighted data secured.

The classical solution runs in $O(n \log(n))$ time for $n$ jobs. It first sorts jobs by end time. Then, for each job $i$, it performs a $O(\log(n))$ binary search to find the latest job that ends before job $i$ starts, known as its predecessor $p(i)$ \cite{kleinberg2005} and accesses this index during the DP recurrence. The per-job binary search is a key bottleneck in the classical solution. 

Our method avoids per-job binary search by precomputing all predecessor indices in a two-pointer $O(n)$ pass after sorting twice. We call our approach \emph{Global Predecessor Indexing (GPI)} which has a time complexity of $O(S(n) + n)$ where $O(S(n))$ is the time complexity of the two sorts done. GPI achieves theoretical $O(n)$ performance when paired with a linear-time sort, and yields strong practical speedups even with general comparison-based sorting. Unlike the classical approach, which performs a binary search per job with poor cache locality, GPI uses a linear scan over sorted data, enabling sequential memory access and hence better cache locality. Since practical sorts like Timsort \cite{timsort2002} are also cache-efficient, the additional sort is a net gain.
 
The final DP step then runs in $O(n)$ time using these predecessor indices. We also discuss the parallelizability of our approach. To our knowledge, this is the first $O(n)$ algorithm for weighted job scheduling when job times admit linear-time sorting, such as bounded integers or smooth distributions.

\section{Background}
\label{sec:background}

\subsection{Classical DP Framework}

Each job $i$ has a start time $s_i$, end time $e_i$, and weight $w_i$. Two jobs $i$ and $k$ are compatible if their intervals do not overlap, i.e., if $e_i \leq s_k$ or $e_k \leq s_i$. All jobs in the chosen subset must be pairwise compatible, and our aim is to find the optimal subset to maximize the sum of its intervals' weights.

The classical solution sorts jobs by increasing end time and computes an array \code{dp}, where \code{dp[i]} stores the maximum achievable total weight using at most the first $i$ jobs \cite{kleinberg2005}. For each job $i$, it finds the latest job $k$ such that $e_k \leq s_i$, defining $p(i) = k$. This is done with an $O(\log(n))$ binary search over the sorted end times. Over all $n$ jobs, this yields $O(n \log(n))$ time complexity. Lastly, we have the DP recurrence:
\[
\code{dp[i]} = \max(\code{dp[i-1]}, w_i + \code{dp[}p(i)\code{]})
\]
where $p(i)$ is the predecessor index of job $i$. If no predecessor exists for a job $i$ (i.e., if no job ends at or before $s_i$), we define $p(i) = 0$. The base case is $\code{dp[0]}=0$, corresponding to an empty job set. \code{dp[n]} stores the final answer. Once $\code{dp}$ is fully computed, the actual set of jobs comprising the optimal solution can be recovered in $O(n)$ time via standard backtracking. Starting from $i = n$, we compare $\code{dp[i]}$ to $\code{dp[i-1]}$: if they are equal, job $i$ was excluded; otherwise, job $i$ was included, and we continue the trace from its predecessor $p(i)$. The selected jobs can be stored in reverse and then returned in sorted order.

\begin{algorithm}[H]
\caption{Classical DP Weighted Job Scheduling}
\begin{algorithmic}[1]
\Require Array \code{jobs} of $n$ jobs
\State Initialize $\code{end_ordered} \gets \code{jobs } \text{sorted by increasing } e_i$
\Function{FindPredecessorBinarySearch}{$i$}
    \State Binary search for largest $k < i$ such that $e_k \leq s_i$
    \State \Return $k$ or $0$ if no such $k$ exists
\EndFunction
\State Initialize $\code{dp} \gets [0, 0, \ldots, 0]$
\For{$i = 1$ to $n$}
    \State $k \gets \textsc{FindPredecessorBinarySearch}(i)$
    \State $\code{dp[i]} \gets \max(\code{dp[i-1]}, w_i + \code{dp[k]})$
\EndFor

\State \Return $\code{dp[n]}$
\end{algorithmic}
\end{algorithm}

%\State Initialize $\code{end_ordered} \gets \code{jobs}$

\subsection{Linear Sorting Algorithms}
\label{sec:sorting}
Standard comparison-based sorting algorithms run in \breaktwo$O(n \log(n))$ time. However, in many practical cases, job times can be sorted in linear time such as when they are bounded-width integers or floats, drawn from uniform distributions, or drawn from a small domain. In such scenarios, specialized linear-time sorting algorithms can achieve $O(n)$ time in the worst case or expectation. We briefly describe several below.

\textbf{Counting Sort \cite{cormen2009}} runs in $O(n + K)$ time, where $K$ is the largest input value. It counts occurrences of each value and reconstructs the sorted array stably. It works best when $K = O(n)$, and is inefficient when the input domain is large or sparse.

\textbf{Radix Sort \cite{cormen2009}} achieves $O(d(n + b))$ time, where $d$ is the number of digits and $b$ is the base. It uses Counting Sort as a stable subroutine per digit. Radix Sort behaves linearly when $d$ is small and can also handle fixed-width decimals.

\textbf{Bucket Sort \cite{aho1983}} divides input into $n$ evenly spaced buckets, sorts each bucket (using a comparison-based sort), and concatenates the results. It has expected $O(n)$ time for roughly uniform inputs, but degrades to $O(n \log(n))$ or worse under skew. It is a good choice when $K \gg n$ or the input range is sparse.

\textbf{Spreadsort \cite{spreadsort2002}} is a hybrid sorting algorithm that combines the linear-time performance of radix sort with the generality of comparison-based sorts. It achieves $O(n)$ in practice by adapting to non-uniform input distributions and degrades to $O(n \log(n))$ only in adversarial cases. It operates on the bit-level representations of numbers, recursively partitioning them by their most significant differing bits. The number of bits used per pass is chosen dynamically to balance recursion depth and bucket size. When buckets become small or indistinct, the algorithm falls back to comparison sort.

\section{Our Approach}
\label{sec:approach}
We present a simple, practical, and efficient algorithm for the weighted job scheduling problem. Our approach eliminates the bottleneck of per-job binary search by introducing a novel preprocessing technique, which we call \textit{Global Predecessor Indexing (GPI)}. After sorting the job intervals twice, GPI computes all predecessor indices in a two-pointer linear pass.

When job times admit linear-time sorting (e.g., bounded-width, uniformly distributed), our full pipeline runs in $O(n)$ time. Even with general comparison-based sorting, GPI significantly outperforms the classical $O(n \log(n))$ solution in practice since per-job binary search is slower than sorting.

In the first phase, we sort the jobs by increasing \textit{end} time and expand each job state to include its ordering within this sort. In the second phase, we sort by increasing \textit{start} time on the array with the expanded jobs. We use both orderings to find all predecessor indices with a clever two-pointer $O(n)$ algorithm in the third phase.

Finally, we use these precomputed predecessor indices with the standard DP from Section~\ref{sec:background} to compute the final numeric answer: the maximum achievable weight from a subset of non-overlapping job intervals. We can also easily find the subset of jobs themselves by backtracking. This eliminates the $O(\log(n))$ per-job binary search overhead with the classical DP, while using the same recurrence. 

\subsection{Sorting Jobs by End Time, Reordering, and Job State Expansion with End-Order}
\label{sec:endtimesort}
We first sort jobs by increasing end time, and ties may be broken arbitrarily. Then, we reorder the given jobs by this sorted order. In other words, $j_1$ is the first job in this order, $j_2$ is next, and so on. We refer to index $i$ as the \enquote{end-order} of $j_i$. By design, $\forall i < n \;:\; e_i \leq e_{i+1}$. As we are reordering, the initial ordering may be discarded.

Lastly, we append each job state $j_i$ with its end-order. 
\begin{align*}
j_i &\gets (s_i, e_i, w_i, i)
\end{align*}
And we now refer to the array $[j_1,\ldots,j_n]$ as the \code{end_ordered} array. For clarity, we also use \enquote{job $i$} to refer to $j_i$, the job with end-order $i$. Additionally, $p(i)$ refers to the end-order of the predecessor job of $j_i$.

\makeatletter
\newif\ifpreprint
\@ifclasswith{elsarticle}{preprint}{\preprinttrue}{\preprintfalse}
\makeatother

\begin{figure}[h]
\centering
\ifpreprint
  \begin{tikzpicture}[scale=1, every node/.style={scale=1}]
\else
  \resizebox{\linewidth}{!}{
  \begin{tikzpicture}[scale=1, every node/.style={scale=1}]
\fi

% Time axis
\draw[->] (0,0) -- (9,0) node[right] {time};

% Intervals: y-level / start / end / job ID
\foreach \y/\start/\endt/\job in {
  1/0/9/4,
  2/0.5/2/1,
  3/3/6/2,
  4/5/6.5/3,
  5/7/9/5
} {
% Interval line
    \draw[very thick] (\start,\y) -- (\endt,\y);
    % Start tick
    \draw[thick] (\start,\y+0.15) -- (\start,\y-0.15);
    % End tick
    \draw[thick] (\endt,\y+0.15) -- (\endt,\y-0.15);
    % Start label (above)
    \node[above=5pt] at (\start,\y) {$s_{\job}$};
    % End label (above)
    \node[above=5pt] at (\endt,\y) {$e_{\job}$};
    % Job ID label (below center)
    \pgfmathsetmacro{\center}{(\start+\endt)/2}
    \node[above=0pt] at (\center,\y) {$j_{\job}$};
}

\ifpreprint
  \end{tikzpicture}
\else
  \end{tikzpicture}
  }
\fi

\caption{Example job intervals reordered by end time.}
\label{fig:intervals-fancy}
\end{figure}
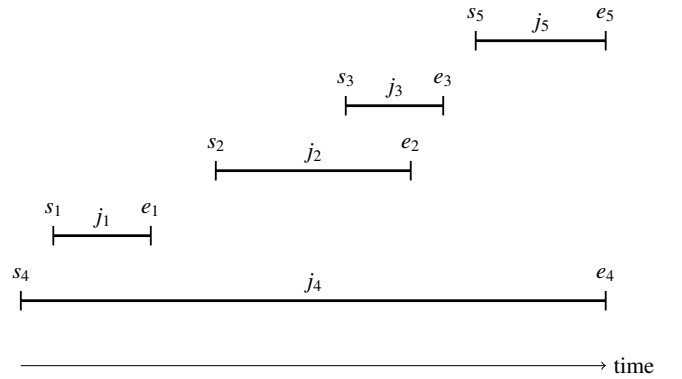
\vspace{-1em}

\subsection{Sorting Jobs by Start Time}
\label{sec:starttimesort}
Unlike classical DP, we sort once again. This time we sort by increasing start time into a new array since we will need both orderings for the next phase. Similarly, we refer to a job's index in the start time ordering as its \enquote{start-order.} By sorting again, we implicitly define a permutation $\pi$ over $\{1, \dots, n\}$ which maps from a job's start-order to its end-order. By design, $\forall i < n \;:\; s_{\pi(i)} \leq s_{\pi(i+1)}$. We refer to $[j_{\pi(1)}, \ldots, j_{\pi(n)}]$ as the \code{start_ordered} array. By construction, $\pi(i)$ is the fourth component of the $i^{\text{th}}$ job in the \code{start_ordered} array\thinspace—\thinspace due to the job state expansion from Section ~\ref{sec:endtimesort}) For clarity, we also use \enquote{job $\pi(i)$} to refer to $j_{\pi(i)}$, the job with start-order $i$.

\begin{table}[h]
\centering
\begin{tabular}{c|c}
start-order $i$ & end-order $\pi(i)$ \\
\hline
1 & 4 \\
2 & 1 \\
3 & 2 \\
4 & 3 \\
5 & 5 \\
\end{tabular}
\caption{Permutation $\pi(i)$ mapping start-order to end-order of the jobs from Figure~\ref{fig:intervals-fancy}, created by sorting again by increasing start time.}
\label{fig:pi-table}
\end{table}

\subsection{Precomputing All Predecessor Indices in $O(n)$ time}
\label{sec:predecessors}

Next, we describe our two-pointer $O(n)$ pass (Algorithm \ref{alg:precompute-predecessors}) to precompute all predecessors, avoiding the per-job binary search from the classical solution. We store predecessor indices in array \code{p}, where \code{p[i]} holds the predecessor $p(i)$ for job $i$.

We find the predecessor for each job in decreasing start-order, i.e., find the predecessor for job $\pi(n)$, then job $\pi(n-1)$, down to job $\pi(1)$. For this, we use \code{start_index} to traverse backward through \code{start_ordered} in our outer loop. We use \code{end_index} to traverse backward through \code{end_ordered} in our inner loop. Before the outer loop begins, we use a single binary search on \code{end_ordered} to initialize \code{end_index} with $p(j_{\pi(n)})$: the predecessor index of job $\pi(n)$. This is the highest end-order of any predecessor job since no job starts later than job $\pi(n)$. Alternatively, we can initialize $\code{end_index}=n$, and let the inner loop decrement $\code{end_index}$ until $p(j_{\pi(n)})$.

Within the outer loop, the inner loop decrements \code{end_index} until we find a job that ends no later than job $\pi(\code{start_index})$ starts. Since we traverse \textit{ backwards} through \code{end_ordered}, the first such job is guaranteed to be $p(j_{\pi(\code{start_index})})$: the highest end-order job which ends before job $\pi(\code{start_index})$ starts. If \code{end_index} hits $0$, then any remaining jobs with unassigned predecessors remain as such since they start before any other job ends. We ensure their entries in $\code{p}$ are $0$. Once the outer loop finishes, all predecessors have been correctly assigned.

\begin{algorithm}[h]
\caption{Precompute Predecessors}
\label{alg:precompute-predecessors}
\begin{algorithmic}[1]
\Require Array \code{end_ordered} of $n$ jobs. Appears as $[j_1, \ldots, j_n]$ where each $(s_i, e_i, w_i, i)$ sorted by increasing $e_i$. 
\Require Array \code{start_ordered} of $n$ jobs. Appears as $[j_{\pi(1)}, \ldots, j_{\pi(n)}]$ where each $(s_{\pi(i)}, e_{\pi(i)}, w_{\pi(i)}, \pi(i))$ sorted by increasing $s_{\pi(i)}$
\Function{EndsNoLaterThanStarts}{$k,i$}
    \State Check if $j_k$ ends no later than $j_i$ starts
    \State \Return \textbf{true} if $e_k \leq s_i$, \textbf{false} otherwise
\EndFunction
\Function{FindPredecessor}{$i$}
    \State Binary search over \code{end_ordered} for largest $k < i$ such that $e_k \leq s_i$
    \State \Return $k$ or $\text{0}$ if no such $k$ exists
\EndFunction

\Function{ComputePredecessors}{$n$}
    \State Initialize $\code{p[}1 \ldots n\code{]} \gets [0 \ldots 0]$  \Comment{predecessor array}
    \State $\code{end_index} \gets \textsc{FindPredecessor}(\pi(n))$
    % \State $\code{p}[\pi(n)] \gets \code{end_index}$
    \For{\code{start_index} $=n$ to $1$}
        \While{$\code{end_index} \geq 1$ \textbf{and} \newline\hspace*{2.724em} (\textbf{not} \textsc{EndsNoLaterThanStarts}($\code{end_index},\newline\hspace*{3em}\pi(\code{start_index})$))}
            \State $\code{end_index} \gets \code{end_index} - 1$
        \EndWhile
        \If{\code{end_index} $=0$}
            \Comment{Remaining jobs with start-order $\leq$ $\code{start_index}$ do not have predecessors, so their entries in \code{p} remain $0$}
            \State \textbf{break} 
        \EndIf
        \State $\code{p[}\pi(\code{start_index})\code{]} \gets \code{end_index}$
        
    \EndFor
    \State \Return $\code{p}$
\EndFunction
\end{algorithmic}
\end{algorithm}

\begin{table}[h]
\centering
\begin{tabular}{c|c}
Job $i$ (end-order $i$) & Predecessor index $p(i)$ \\
\hline
1 & 0 \\
2 & 1 \\
3 & 1 \\
4 & 0 \\
5 & 3 \\
\end{tabular}
\caption{Predecessor indices $p(i)$ corresponding to the intervals in Figure~\ref{fig:intervals-fancy}.}
\label{fig:pred-table}
\end{table}

\subsubsection*{Time Complexity}

We analyze the time complexity of Algorithm ~\ref{alg:precompute-predecessors}: Precompute Predecessors. First, we perform a single binary search to find $p(\pi(n))$ which takes $O(\log(n))$ time. Next, although the \code{while} loop is nested within the \code{for} loop, the crucial observations are that:
\begin{itemize}
    \item \code{start_index} decrements monotonically and visits each index from $[1, \ldots, n]$ at most once, and so the number of iterations of the outer loop is at most $n$.
    \item \code{end_index} is initialized before the outer loop, decrements monotonically, and visits each index from $[0, \ldots, n]$ at most once. The \textit{total} number of iterations of the inner loop across the function is at most $n$.
\end{itemize}

Each iteration of the inner loop performs a fixed number of $O(1)$ operations: evaluating \textsc{EndsNoLaterThanStarts} (two array accesses and a comparison) and decrementing \code{end_index}. Each iteration of the outer loop also performs a fixed number of $O(1)$ operations: evaluating the comparison on line 16, the array assignment on line 19, and decrementing \code{start_index}. All of these operations\thinspace—\thinspace including arithmetic updates, logical comparisons, and array accesses or assignments\thinspace—\thinspace are primitive and take constant time, i.e., $O(1)$. Since the total number of \code{start_index} and \code{end_index} decrements is bounded by $2n$, and each decrement involves only $O(1)$ work, the total time complexity is $O(n)$.

\subsubsection*{Correctness} 
We prove correctness of our novel preprocessing step via a Loop Invariant over the outer ($\textbf{for}$) loop on line 12.

\paragraph{Loop Invariant}
At the beginning of line 16 during iteration \code{start_index} $=i$, \code{end_index} is the predecessor of job \breaktwo $\pi(\code{start_index}) = i$. Although the loop begins at line 12, this invariant is defined to hold specifically at line 16, just before the predecessor for job $\pi(\code{start_index})$ is assigned on line 19.

\paragraph{Base Case}
Before the first iteration, \textsc{FindPredecessor} uses binary search to initialize \code{end_index} with $p(j_{\pi(n)})$. Since \breaktwo$\code{start_index}=n$ in the first iteration and by the correctness of the classical solution, the loop invariant holds true. The inner loop will not decrement \code{end_index} during the first iteration.

\paragraph{Maintenance}
Assume the invariant holds at the start of line 16 during iteration $\code{start_index}=i$. Hence, $\code{end_index}$ equals $p(\pi(\code{start_index}))$. Now, we show that the inviarant also holds true in the next iteration where \code{start_index} $=(i-1)$. First, we prove that $p(\pi(i-1)) \leq p(\pi(i))$:
\begin{proof}
By way of contradiction, assume the opposite, so $p(\pi(i-1)) > p(\pi(i))$. By definition of predecessor, job $p(\pi(i-1))$ ends no later than job $\pi(i-1)$ starts. By construction, job $\pi(i-1)$ starts no later than job $\pi(i)$ starts. Then, by transitivity, job $p(\pi(i-1))$ ends no later than job $\pi(i)$ starts. 

Due to our assumption for contradiction that $p(\pi(i-1)) > p(\pi(i))$, $p(\pi(i))$ can not be the predecessor for job $\pi(i)$ since there is a higher end-order job, job $p(\pi(i-1))$, that ends no later than job $\pi(i)$ starts. However, $p(\pi(i))$ is indeed defined to be the predecessor for job $\pi(i)$.
\end{proof}

Now, we know that the correct \code{end_index} for iteration \code{start_index} $=(i-1)$ occurs at or before the correct $\code{end_index}$ from iteration $\code{start_index}=i$. As we are simply doing a linear search backwards through \code{end_ordered}, we are guaranteed to find the job with the highest end-order that ends no later than job $\pi(i-1)$ starts: its predecessor. Lines 13-15 decrement $\code{end_index}$ until the invariant is restored or maintained, so that when control reaches line 16, the invariant still holds.

\paragraph{Termination}
When the loop terminates after iteration\breaktwo\code{start_index} $=1$, the invariant has held at line 16 in every iteration \code{start_index} $=i$ that was reached and correctly assigning $p[\pi(i)]$ to \code{end_index} on line 19. Any outer-loop iterations that were not reached were because \code{end_index} hit $0$, which all entries in \code{p} were initialized to anyway. Since $\pi$ is a permutation over $\set{1, \ldots, n}$ and $\code{start_index}$ looped from $n$ to $1$, then $\code{p[i]}$ was set at most once for each $i \in \set{1, \ldots, n}$ with the correct \code{end_index}. Thus, $\code{p[i]}$ correctly stores the predecessor for all jobs $i$.

\subsubsection*{Parallelizability}
The outer loop can be split across multiple threads. Each thread can handle a continuous segment of \code{start_index} values in reverse order, say $[u_m, \ldots, u_1]$. The cost is one binary search per thread to initialize \code{end_index} with the predecessor of job $u_m$. As each thread can compute the predecessor of many jobs, this is a good tradeoff.

\subsection{Dynamic Programming}
Assuming the predecessor indices $p(i)$ have already been computed (see Section~\ref{sec:predecessors}) and we have GPI's resulting predecessor lookup table \code{p}, we now describe the final DP step from Section~\ref{sec:background}:
\[
\code{dp[i]} = \max(\code{dp[i-1]}, w_i + \code{dp[p[i]]})
\]
with base case \code{dp[0]} $=0$ by convention. During the DP, we can retrieve each job's predecessor in constant time via \code{p}, eliminating the need for per-job binary search and providing the main speedup of this paper. Since all array accesses and maximum computations are $O(1)$, the total time to compute the DP table, the final numeric answer, and the optimal subset of jobs themselves is $O(n)$.

\subsubsection*{Correctness}

As in the classical solution, \code{dp[i]} holds the maximum achievable total weight among the first $i$ jobs. We prove correctness by induction on $i$. The base case \code{dp[0]} $=0$ is valid as no jobs have been selected. Assume $\code{dp[i-1]}$ is correct. Then $\code{dp[i]}$ either inherits the optimal solution for the first $(i-1)$ jobs or takes the optimal solution for the predecessor $p(i)$\thinspace---\thinspace\code{dp[p[i]]})\thinspace---\thinspace plus $w_i$. Since $p(i)$ was correctly computed, and since the recurrence takes the maximum of these two cases, the solution remains optimal. Since $\code{dp[n]}$ now represents the maximum achievable total weight among the first $n$ jobs, and there are $n$ jobs total, it holds the final answer after the loop ends.

\subsubsection*{Parallelizability}
Since all predecessor relationships are known in advance, the DP step can be parallelized using Kahn's topological sort \cite{kahn1962}, where each job depends only on its predecessor job's DP value. The effectiveness of this parallelization depends on how often jobs are predecessors to multiple other jobs. 

\section{Experimental Results}
\label{sec:experiments}

We benchmark GPI with linear sorting and GPI with Timsort against the classical $O(n \log(n))$ solution. Although GPI achieves theoretical $O(n)$ with linear sorting, GPI with Timsort offers the best practical performance due to Timsort's optimized design. We release our implementation and benchmark code.\footnote{\url{https://github.com/amitjoshi2724/gpi-job-scheduling}}

\subsection{Benchmark Design}

We design four controlled experiments to evaluate GPI performance under different job generation patterns, each capturing distinct structural characteristics of real-world workloads. Both classical and linear-time algorithms are implemented in Python 3 and benchmarked on identical hardware with cross-validated results.
Runtimes are averaged over 10 trials per input size to mitigate noise, and garbage collection triggered only before trials. For each experiment, we vary $n$ from $1{\,}000$ to $100{\,}000$ jobs and report two metrics. The first metric is \textbf{Overall runtime (seconds)}, which measures wall-clock time to solve a complete instance. The second metric is \textbf{per-job runtime (seconds/job)} which is computed as total time divided by~$n$ and is useful for visualizing marginal cost and highlighting logarithmic overhead within the classical solution.

\subsection{Experiment Details}

\paragraph{Experiment 1: Random Integer Times}
Job start and end times are drawn uniformly from $[0, 10^6]$, representing generic conditions with arbitrary durations, random overlaps, and diverse predecessor relationships. This baseline test resembles mixed workloads in general-purpose scheduling applications. GPI Linear uses Radix Sort as job times are bounded-width integers.
\vspace{-1em}
\begin{figure}[H]
    \centering
    \includegraphics[width=\linewidth]{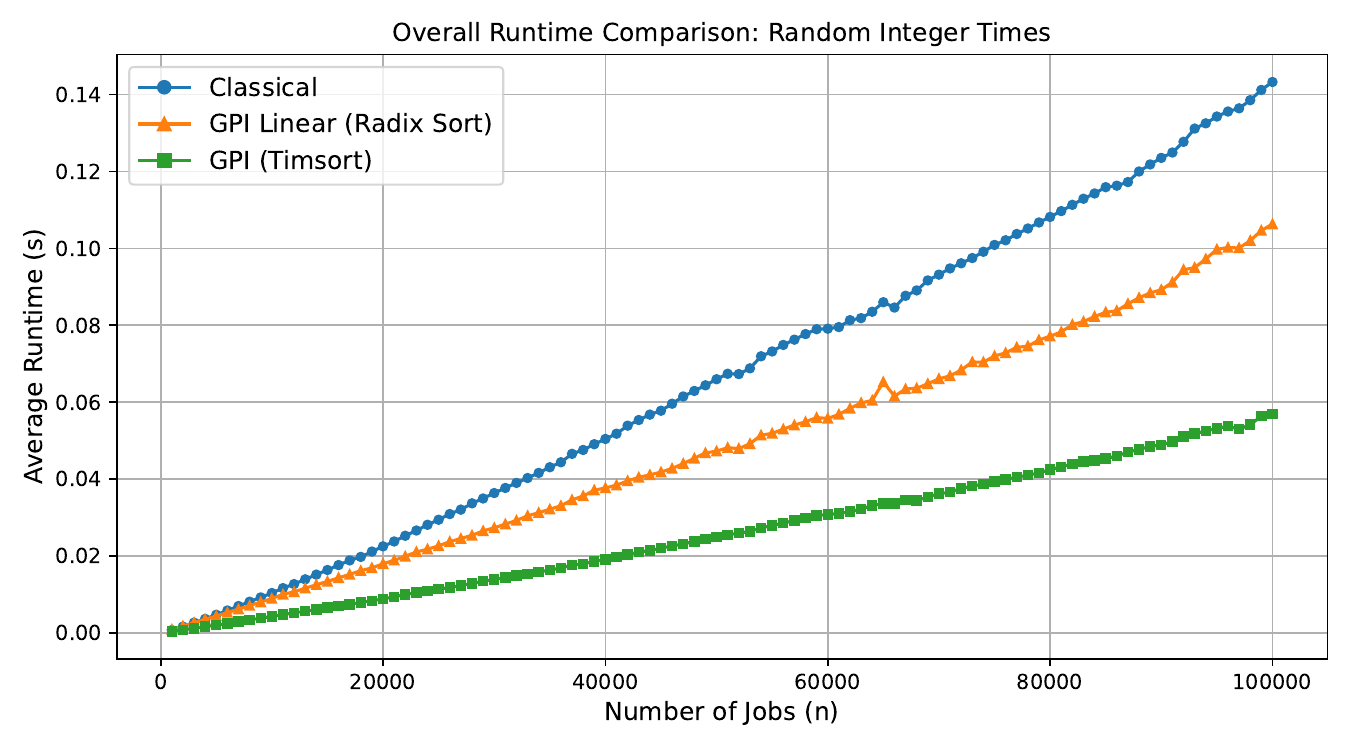}
    \label{fig:exp1-runtime-total}
\end{figure}
\vspace{-3em}
\begin{figure}[H]
    \centering
    \includegraphics[width=\linewidth]{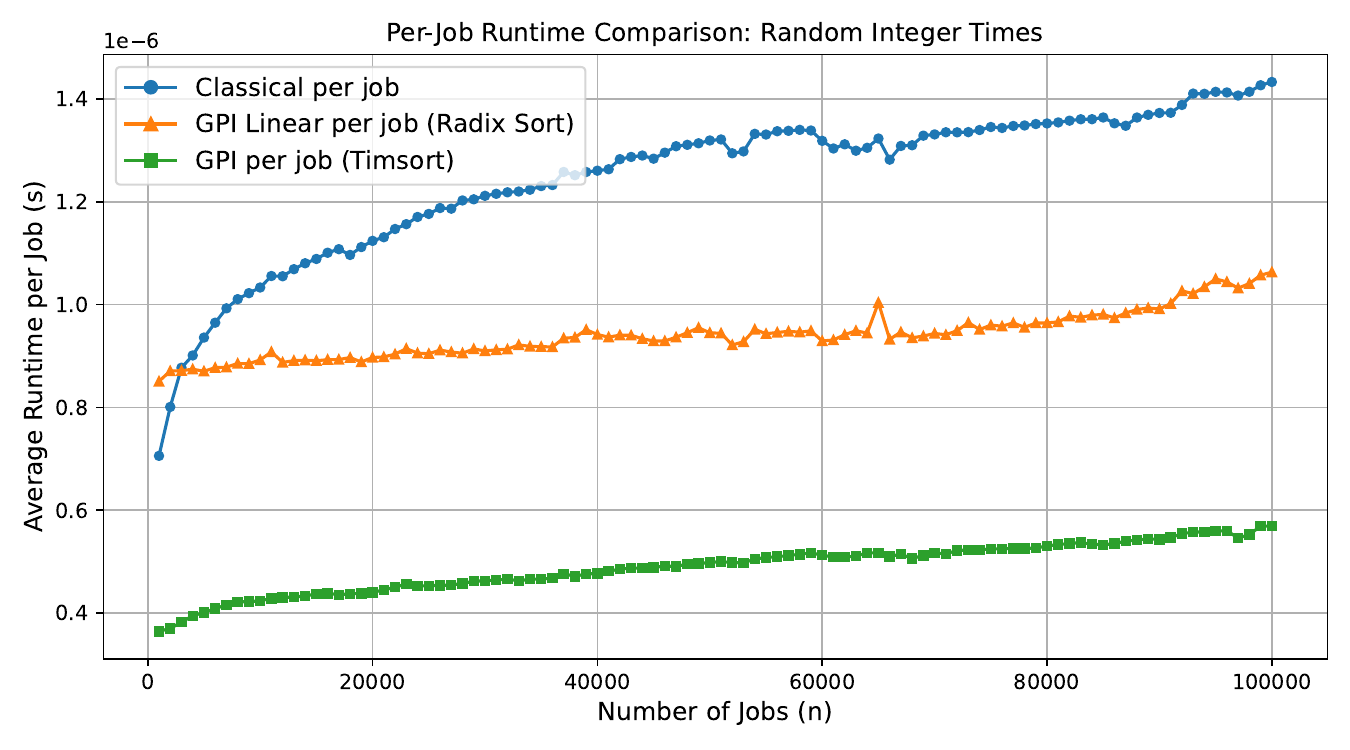}
    \label{fig:exp1-runtime-per-job}
\end{figure}
\vspace{-2em}
\paragraph{Experiment 2: Normally Distributed Start Times}
Job start times are drawn from a truncated normal distribution $N\left(\frac{K}{2}, \left(\frac{K}{10}\right)^2\right)$ within $[0, K=10^9]$, and float durations are sampled uniformly from $[1, 10^6]$. This models job bursts in real-world systems with smooth, dense clusters. GPI Linear uses Spreadsort, which efficiently handles such non-uniform input distributions.
\vspace{-1em}
\begin{figure}[H]
    \centering
    \includegraphics[width=1\linewidth]{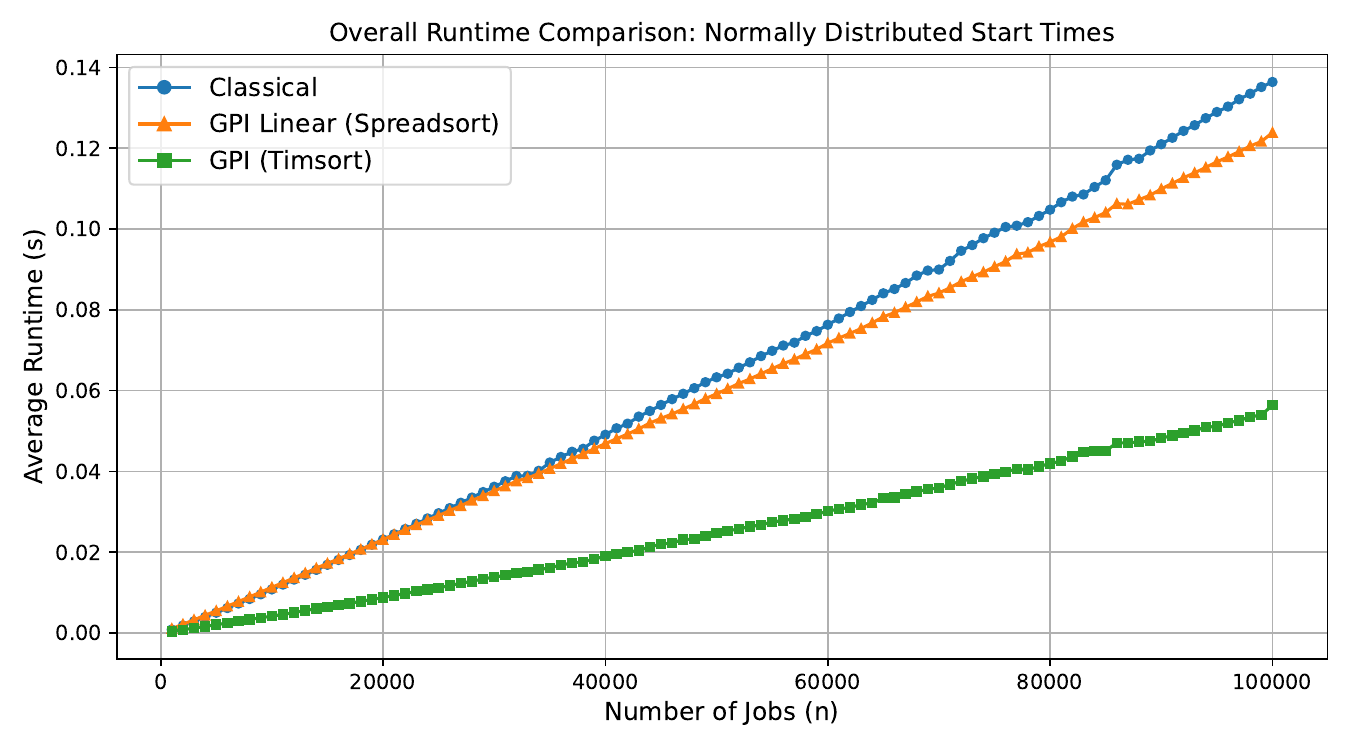}
    \label{fig:exp2-runtime-total}
\end{figure}
\vspace{-2.5em}
\begin{figure}[H]
    \includegraphics[width=1\linewidth]{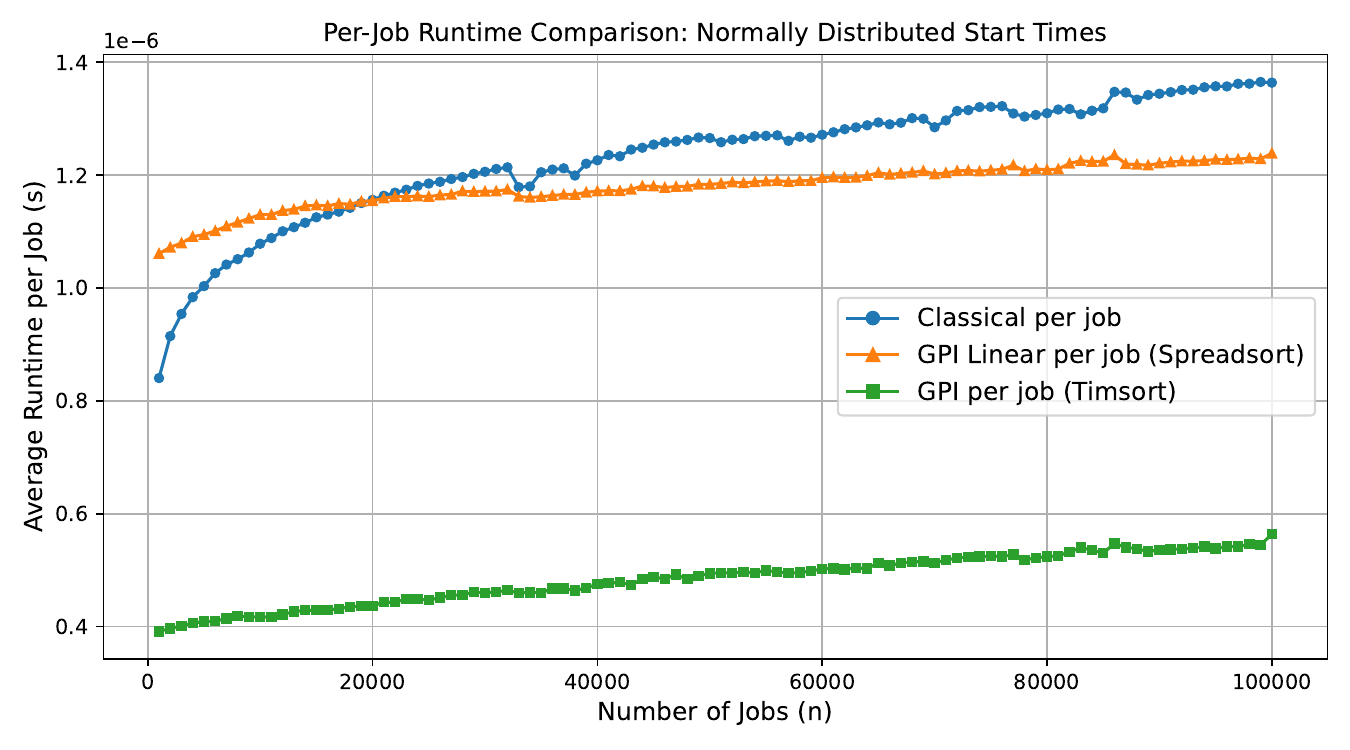}
    \label{fig:exp2-runtime-per-job}
\end{figure}
\vspace{-2em}

\paragraph{Experiment 3: Zipf Durations with Early Start Bursts}
Job start times follow a truncated exponential distribution with scale $\frac{K}{10}$ (clustered near 0), modeling early start bursts in real-world workloads. Durations follow a scaled Zipf distribution with exponent $s = 2$ (i.e., $d = 100 \cdot Z$ for $Z \sim \text{Zipf}(2)$), capped at $10^6$. This yields mostly short jobs with few long ones, causing some jobs to have nearby predecessors while others have distant ones. This disrupts cache locality during predecessor lookups during the DP. GPI Linear uses Spreadsort for this heavily skewed input distribution.
\vspace{-1em}
\begin{figure}[H]
    \centering
    \includegraphics[width=1\linewidth]{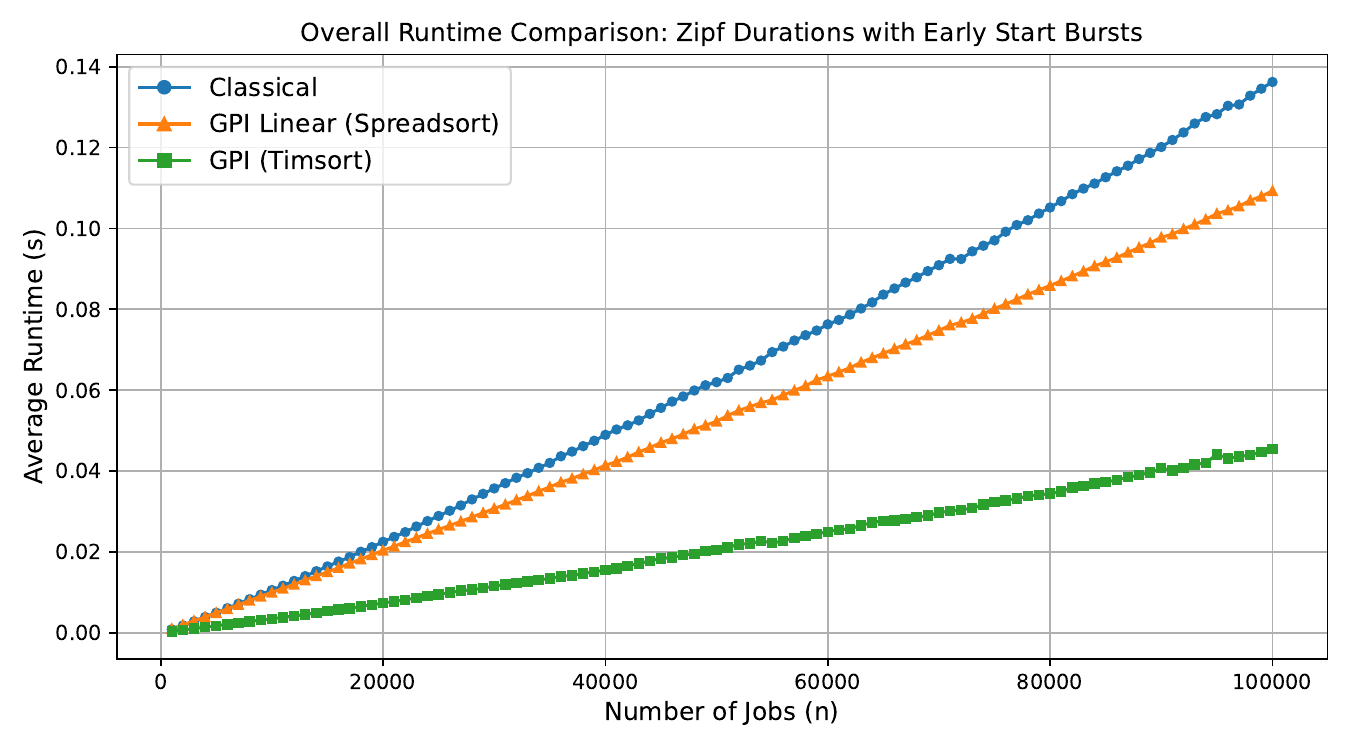}
    \label{fig:exp3-runtime-total}
\end{figure}
\vspace{-2.5em}
\begin{figure}[H]
    \includegraphics[width=1\linewidth]{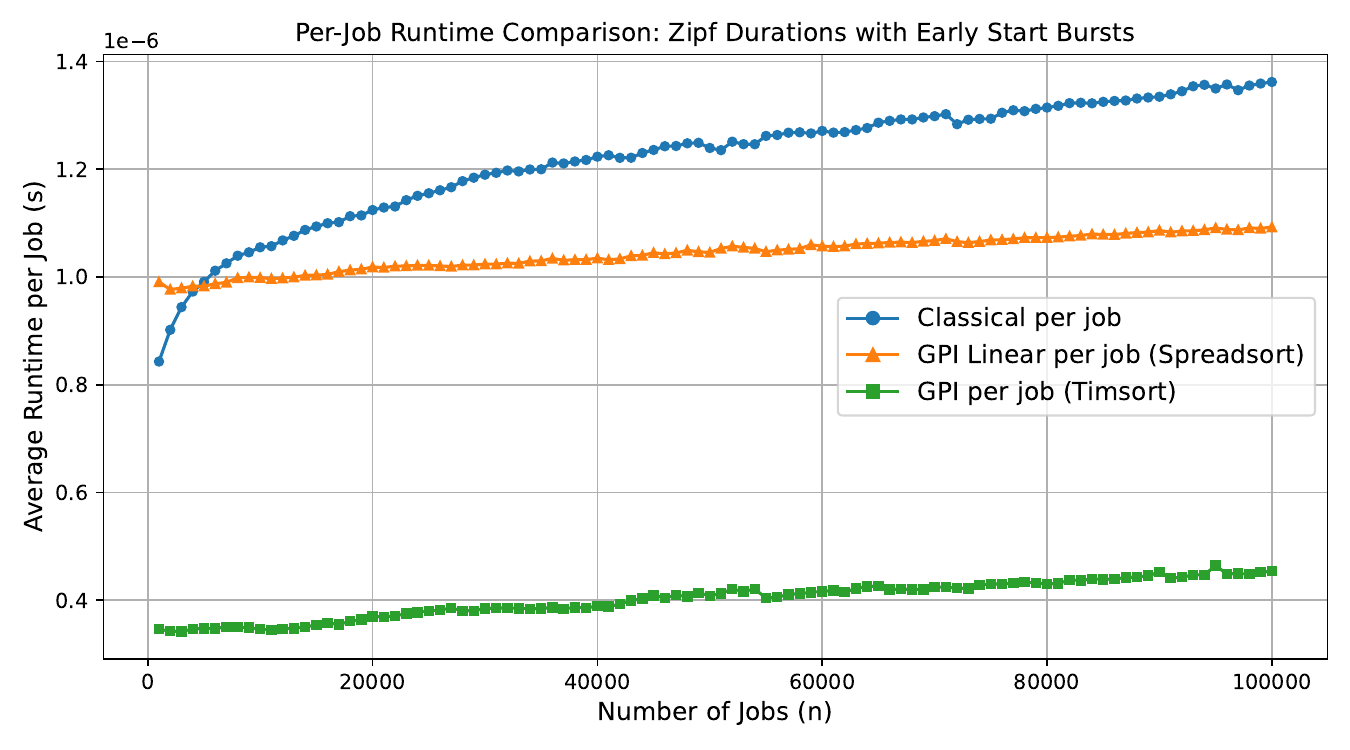}
    \label{fig:exp3-runtime-per-job}
\end{figure}
\vspace{-2em}

\paragraph{Experiment 4: Bucket-Sort-Friendly Uniform Start Times}
Job start times are drawn uniformly from $[0, K=10^9]$ so $K \gg n$, and float durations are sampled uniformly from $[1, 10^6]$, so the end times are approximately uniform as well. This produces a well-spread timeline ideal for Bucket Sort, which our GPI Linear uses. The uniform distribution ensures minimal collisions per bucket. Collisions are sorted using Timsort (a hybrid merge/insertion sort that exploits ordered subarrays ~\cite{timsort2002}), highlighting Bucket Sort's strengths when input times are approximately uniformly distributed across a wide domain.
\vspace{-1em}
\begin{figure}[H]
    \centering
    \includegraphics[width=\linewidth]{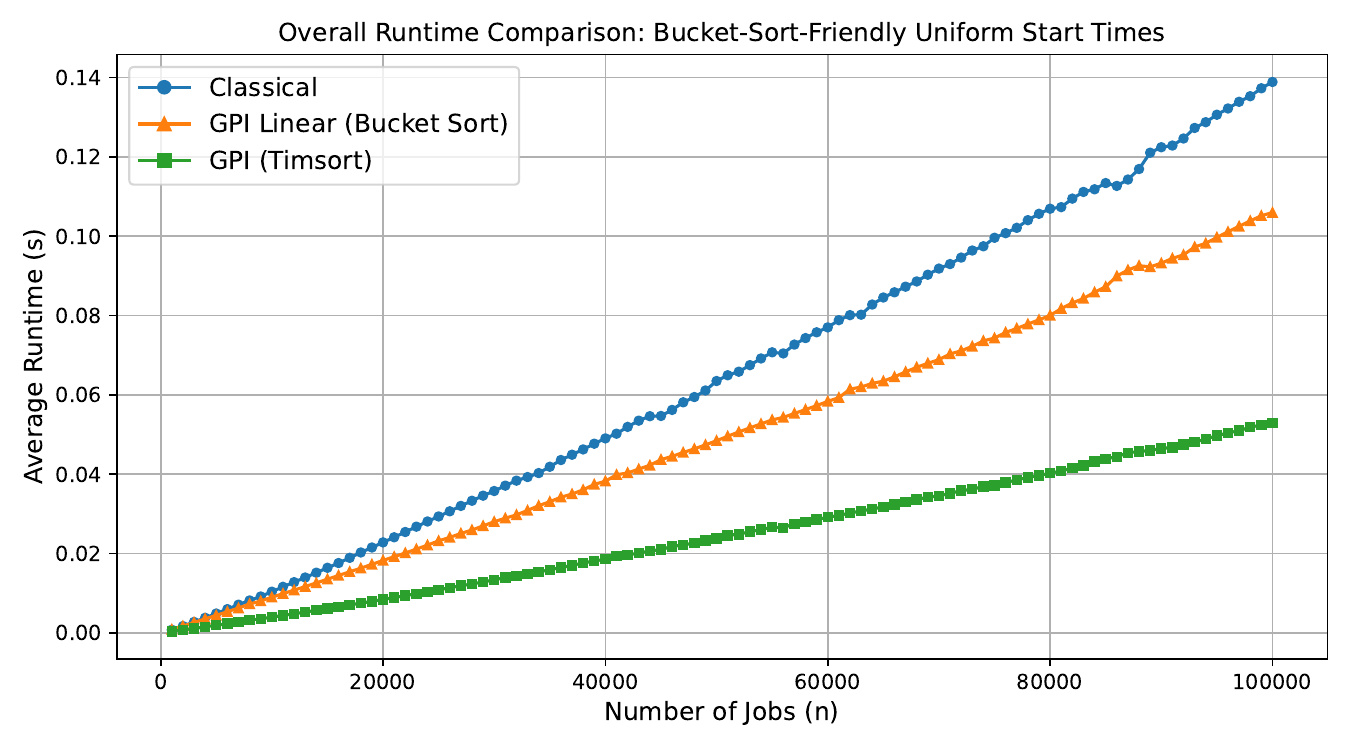}
    \label{fig:exp4-runtime-total}
\end{figure}
\vspace{-2.5em}
\begin{figure}[H]
    \includegraphics[width=\linewidth]{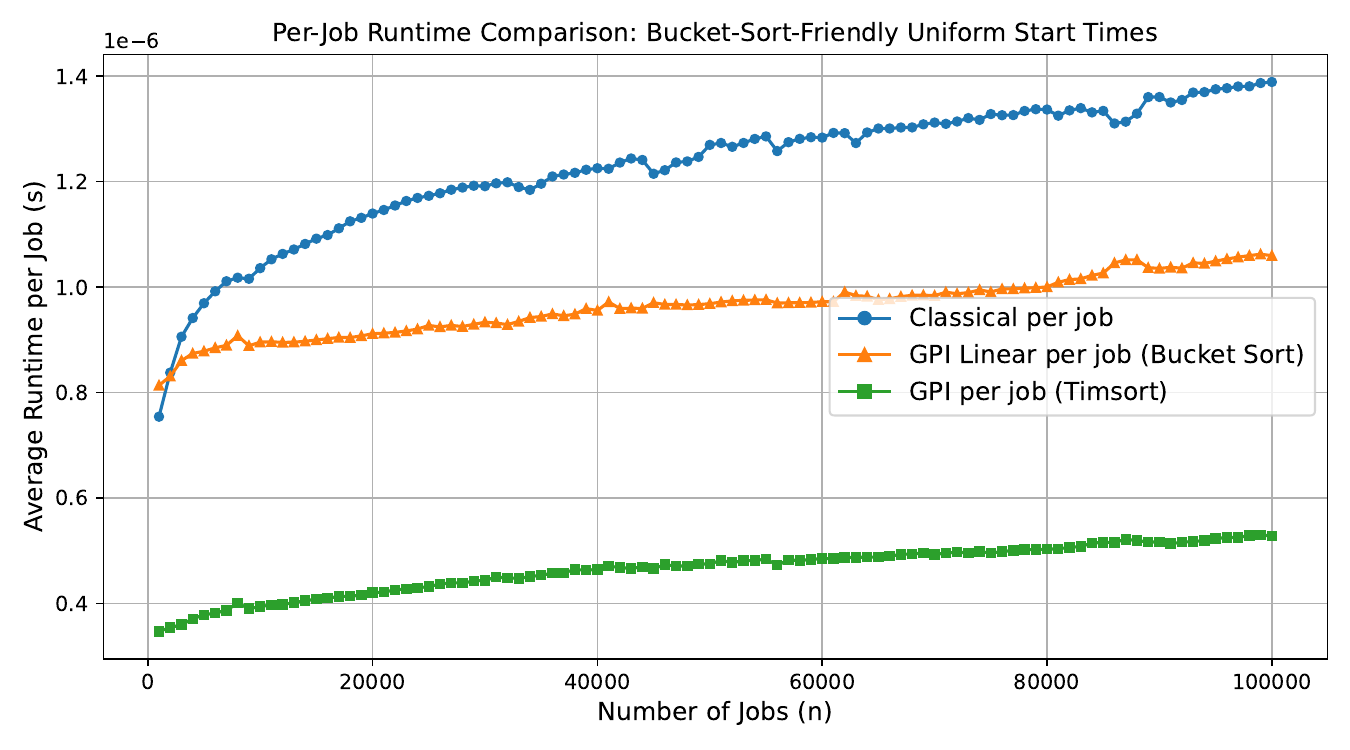}
    \label{fig:exp4-runtime-per-job}
\end{figure}
\vspace{-3em}
\subsection{Discussion}

Our results align with theoretical expectations. The classical solution always shows mild upward curvature in the Overall Runtime charts, consistent with its $O(n \log(n))$ time complexity. In the Per-Job Runtime charts, Classical DP displays a clear logarithmic trend, reflecting the increasing marginal cost of processing each additional job.

GPI Linear in Experiments 1 and 4 achieves true linear scaling, producing linear slopes in Overall Runtime charts with flat Per-Job Runtime curves, confirming constant marginal cost per job. While GPI Linear shows a slight upward slant in per-job runtime, this reflects system-level overhead (possibly due to cache misses) rather than algorithmic inefficiency. The scaling outperforms the classical solution with increasing $n$ and supports our linear-time claim.

GPI Linear in Experiments 2 and 3 uses Spreadsort on non-uniform input distributions and achieves good speedups over the classical solution despite Python-C++ conversion overhead via Pybind11. While speedups are less pronounced than in Experiments 1 and 4, performance improves with larger $n$ and Per-Job Runtime curves remain relatively flat, demonstrating near-linear runtime and near-constant marginal costs even under heavily skewed, non-ideal conditions.

GPI with Timsort consistently outperforms all methods by eliminating the per-job binary search bottleneck. Despite a very slight logarithmic curve in per-job runtime, overall performance remains near-linear. At $n = 100{,}000$, GPI Timsort runs 2-3 times faster than the classical solution\thinspace—\thinspace an overwhelming practical speedup despite identical $O(n \log(n))$ theoretical worst-case complexity.

\section{Conclusion}
\label{sec:conclusion}

We have shown that GPI significantly accelerates WJS in the general case by avoiding binary search, and proven our algorithm's correctness and runtime of $O(S(n) + n)$. When job times permit linear-time sorting, GPI solves WJS in $O(n)$ time, beating the classical solution's $O(n \log(n))$. Even with comparison-based sorting, GPI greatly outperforms the classical solution in practice, and outperforms GPI paired with linear-time sorts. GPI sorts twice, and then replaces per-job binary searches with two linear-time passes to find all predecessors and uses array-indexed predecessor lookups during DP. Our approach gives a theoretical and practical speedup over the classical solution as well as being conceptually simple and parallelizable. Its cache-friendly linear scan avoids the poor locality of per-job binary search, contributing to its 2–3× real-world speedup. We hope that our technique of sorting by one key, expanding states to store this sort order, and then sorting by another key will be used to solve other problems as well.

\bibliographystyle{elsarticle-num}  % or use 'elsarticle-num' if submitting to IPL
\bibliography{references}

\end{document}